\documentclass[3p,times]{elsarticle}

\usepackage{ecrc}

\usepackage{epstopdf}

\volume{00}

\firstpage{1}

\journalname{Nuclear Physics A}

\runauth{Author1 et al.}

\jid{nupha}

\jnltitlelogo{Nuclear Physics A}

\usepackage{graphicx}
\usepackage{amsmath,amssymb}
\begin{document}

\begin{frontmatter}



\title{Phenomenology of photon and di-lepton production in relativistic nuclear collisions}

\author{Elena Bratkovskaya}
\address{Institute for Theoretical Physics and Frankfurt Institute for Advanced Studies, Johann Wolfgang Goethe Universit\"at, Frankfurt am Main, Germany}




\begin{abstract}
We discuss the latest theoretical results on direct photon and dilepton
production from relativistic heavy-ion collisions.
While the  dilepton spectra at low invariant mass show  in-medium effects like a 
collisional broadening of the vector meson spectral functions, the dilepton yield 
at high invariant masses (above 1.1 GeV) \ is dominated by QGP contributions 
for central heavy-ion collisions at relativistic energies.
The present status of the photon $v_2$ "puzzle" --
a large elliptic flow $v_2$ of the direct photons experimentally 
observed at RHIC and LHC energies -  is also addressed.
The role of hadronic and partonic sources for the photon spectra 
and $v_2$ is considered as well as the possibility to 
subtract the QGP signal from the experimental observables.
\end{abstract}


\end{frontmatter}



\section{Introduction}
\label{intro}
Electromagnetic probes - real photons and di-lepton pairs - are one of the most
promising probes of the Quark-Gluon-Plasma (QGP) formed in ultra-relativistic
collisions of heavy ions  \cite{FeinbShur}. They also provide information about the modification of hadron properties in the dense and hot hadronic medium 
which can shed some light on chiral symmetry restoration (cf. \cite{ChSym} 
and references therein).
The major advantages are related to the fact that dileptons and real 
photons are emitted from different stages of the reaction and are
very little effected by final-state interactions;
they provide almost undistorted information about their production channels.
However, there are disadvantages, too, due to the low emission rate;
the production from the hadronic corona as well as the existance of
many production sources, which cannot be individually disentangled by 
experimental data.
For an experimental review on the dilepton and photon measurements at 
SPS, RHIC and LHC energies see Ref. \cite{RuanQM14}.

Since dileptons and photons are emitted over the entire history of the heavy-ion
collision, from the initial hard nucleon-nucleon scatterings through the
hot and dense (partonic) phase and to the hadron decays after
freeze-out, it is very important to model properly the full time evolution 
of heavy-ion collisions. For that the dynamical models - microscopic 
transport or hydrodynamical approaches - have to be applied for
disentangling the various sources that contribute to the final
observables measured in experiments.

\section{Modeling of photon/dilepton emission rates}

\noindent
I. The equilibrium emission rate of electromagnetic probes from 
thermal field theory can be expressed as \cite{Rate1,Rate2}: \\
1) for photons with 4-momentum $q=(q_0,\vec q)$:
\begin{eqnarray}
q_0 {d^3R\over d^3q}= -{\dfrac{g_{\mu\nu}}{(2\pi)^3}} 
	Im \Pi^{\mu\nu} (q_0=|\vec{q}|) f(q_0,T);
\label{RatePh}	
\end{eqnarray}
2) for dilepton pairs with 4-momentum  $q=(q_0,\vec q)$, where 
$q=p_+ + p_-$ and $p_+=(E_+,\vec p_+), p_-=(E_-,\vec p_-)$:
\begin{eqnarray}
E_+E_- {d^3R \over {d^3p_+d^3p_-}}= \dfrac{2e^2}{(2\pi)^6} \dfrac{1}{q^4}
	L_{\mu\nu} Im \Pi^{\mu\nu} (q_0,|\vec{q}|)f(q_0,T).
\label{RateDil}	
\end{eqnarray}
Here the Bose distribution function is $f(q_0,T) =1/(e^{q_0/T}-1)$;
$L_{\mu\nu}$ is the electromagnetic leptonic tensor,
$\Pi^{\mu\nu}$ is the retarded photon self-energy at finite temperature $T$
related to the electromagnetic current correlator
$\Pi^{\mu\nu} \sim i\int d^4x e^{ipx}\left\langle[J_\mu(x),J_\nu(0)]\right\rangle|_T $.
Using the Vector-Dominance-Model (VDM) $Im\Pi^{\mu\nu}$ can be related to the 
in-medium $\rho$-meson spectral function from many-body approaches \cite{RappWam97}
which, thus, can be probed by dilepton measurements directly.
The photon rates for $q_0\to 0$ are related to the electric conductivity $\sigma_0$
which allows to probe the electric properties of the QGP \cite{Cass13PRL}.
We note that Eqs.(\ref{RatePh}),(\ref{RateDil}) are applicable for systems 
in thermal equilibrium, whereas the dynamics of heavy-ion collisions is 
generally of non-equilibrium nature.

\noindent
II. The non-equilibrium emission rate from relativistic kinetic theory 
\cite{Rate2,RateKT}, e.g. for the process $1+2\to \gamma +3$, is
\begin{equation}
q_0 {d^3R\over d^3q}= \int \dfrac{d^3p_1}{2(2\pi)^3 E_1} 
\dfrac{d^3p_2}{2(2\pi)^3 E_2} \dfrac{d^3p_3}{2(2\pi)^3 E_3} \
(2\pi)^4 \ \delta^4(p_1+p_2-p_3-q) \ |M_{if}|^2 \ 
\dfrac{f(E_1) f(E_2) (1\pm f(E_3))}{2(2\pi)^3},
\label{RateKT}
\end{equation}
where $f(E_i)$ is the distribution function of $i$-particle ($i=1,2,3$),
which can be hadrons (mesons and baryons) or partons.
$M_{if}$ is the matrix element of the reaction which has to be evaluated 
on a microscopical level. For hadronic interactions One-Boson-Exchange
models or chiral models are used to evaluate $M_{if}$  on the level 
of Born-type diagrams. However, for a consistent consideration 
of such elementary process in the dense and hot hadronic environment, 
it is important to account for the in-medium modification of hadronic
properties \cite{RappWam97}, i.e. many-body 
approaches such as self-consistent $G$-matrix calculations have to be applied.

\section{Photons}
\subsection{Photon production sources}

There are different production sources of photons in $p+p$ and $A+A$ collisions:\\
1) {\it Decay photons } --  most of the photons seen in $p+p$ 
and $A+A$ collisions stem from the hadronic decays: \\ 
$m \to \gamma + X, \ \  m = \pi^0, \eta, \omega, \eta^\prime, a_1, ...$. \\
2) {\it Direct photons} -- obtained by subtraction of the decay 
photon contributions from  the inclusive (total) spectra measured
 experimentally.\\
$(i)$ 
The are a few sources of direct photons at large transverse momentum $p_T$,
so called {\it 'hard'} photons:
the 'prompt' production from the initial hard $N+N$ collisions and the 
photons from the jet fragmentations, which are the standard pQCD type of
processes. 
The latter, however, might be modified in $A+A$ contrary to $p+p$ 
due to the parton energy loss in the medium.
In $A+A$ collisions at large $p_T$ there are also photons  form the 
jet-$\gamma$-conversion in the QGP and jet-medium photons from the scattering of hard partons
with thermalized partons ($q_{hard}+q(g)_{QGP} \to \gamma + q(g)$. \\
$(ii)$ 
At low $p_T$ the photons come from the thermalized QGP 
as well as from {\it hadronic} interactions: \\
$\bullet$ 
The {\it 'thermal'} photons from the QGP arise mainly from $q\bar q$
annihilation  ($q+\bar q \to g +\gamma$) and Compton scattering 
($q(\bar q) + g \to q(\bar q) + \gamma$)
which can be calculated in leading order pQCD  \cite{AMY01}.
However, the next-to-leading order corrections turn out to be 
also important \cite{JacopoQM14}.\\
$\bullet$
{\it Hadronic} sources of photons are related to \\
1) secondary mesonic interactions as 
$\pi + \pi \to \rho + \gamma, \ \rho + \pi \to \pi + \gamma, \
\pi + K \to \rho + \gamma, ....$ 
The binary channels with $\pi, \rho$ have been evaluated in effective 
field theory \cite{Kapusta91} and used in transport 
model calculations \cite{HSD08,Linnyk:Photon} within the extension
for the off-shellness of $\rho$-mesons due to the broad spectral function. Alternatively, the binary hadron rates (\ref{RateKT}) have been derived 
in the massive Yang-Milles approach in Ref. \cite{TRG04} and been
often used in hydro calculations. \\
2) hadronic bremsstrahlung, such as meson-meson ($mm$)
and meson-baryon ($mB$) bremsstrahlung
$m_1+m_2\to m_1+m_2+\gamma, \ \ m+B\to m+B+\gamma$, where 
$m=\pi,\eta,\rho,\omega,K,K^*,...$ and $B=p,\Delta, ...$. 
Here the leading contribution corresponds to the radiation 
from one charged hadron scattered with the neutral hadron.

\subsection{Direct photons: the $v_2$ 'puzzle'}

The photon production has been measured early in relativistic heavy-ion
collisions  by the WA98 Collaboration in S+Au and Pb+Pb collisions 
at SPS energies  \cite{WA98}.
The model comparisons with experimental data shows that the high 
$p_T$ spectra are dominated by the hard 'prompt' photon production 
whereas the 'soft' low $p_T$ spectra stem from hadronic sources 
since the thermal QGP radiation  at SPS energies is not large. 
Moreover, the role of hadronic bremsstrahlung turns out to be very important 
for a consistent description of the low $p_T$ data as has been found in 
expanding fireball model calculations \cite{LiuRapp07} 
and in the HSD (Hadron-String-Dynamics) transport approach \cite{HSD08}.  
Unfortunately, the accuracy of the experimental data at low $p_T$ 
did not allow to draw further solid conclusions.

The measurement of photon spectra by the PHENIX Collaboration 
\cite{PHENIX2010} stimulated a new wave of interest for 
direct photons from the theoretical side since at RHIC energies 
the thermal QGP photons have been expected to dominate the spectra. 
A variety of model calculations based on fireball, Bjorken hydrodynamics, 
ideal hydrodynamics with different initial conditions and 
Equations-of-State (EoS) turned out to show substantial differences 
in the slope and magnitude of the photon spectra 
(for the model comparison see Fig. 47 of \cite{PHENIX2010}
and corresponding references therein).

The recent observation by the PHENIX Collaboration \cite{PHENIX1}
that the elliptic flow $v_2(p_T) $ of 'direct photons' produced in
minimal bias Au+Au collisions at $\sqrt{s_{NN}}=200$~GeV is
comparable to that of the produced pions was a surprise and in
contrast to the theoretical expectations and predictions 
\cite{Chatterjee:2005de,Liu:2009kq,Dion:2011vd,Chatterjee:2013naa}.
Indeed, the photons produced by partonic interactions in the
quark-gluon plasma phase have not been expected to show a considerable
flow because - in a hydrodynamical picture - they are dominated by
the emission at high temperatures, i.e. in the initial phase before
the elliptic flow fully develops.
Since the direct photon $v_2(\gamma^{dir})$ is a 'weighted average' 
by the corresponding yields $N_i$ of the elliptic flow of 
individual sources $i$:
$v_2 (\gamma^{dir}) =  \frac{\sum _i  v_2 (\gamma^{i}) N_i }{\sum_i N_i }.$
Thus, a large QGP contribution gives smaller $v_2(\gamma^{dir})$.

A sizable photon $v_2$ has been observed also by the ALICE Collaboration
\cite{ALICE_v2}. None of the theoretical models could describe simultaneously 
the photon spectra and $v_2$ which may be noted as a 'puzzle' for theory.
Moreover, the PHENIX and ALICE Collaborations have reported recently 
the observation of non-zero triangular flow $v_3$ (see  \cite{RuanQM14,BockQM14}).
Thus, the consistent description of the photon experimental data
remains as a challenge for theory and has stimulated new ideas and 
developments. Some of them we  briefly discuss in the  Section 3.3.

\subsection{Towards the solution of the photon $v_2$ 'puzzle': theory}

\subsubsection{Developments in hydrodynamical models}

I.) The first hydrodynamical calculations on photon spectra 
were based on the ideal hydro with smooth Glauber-type initial 
conditions (cf. \cite{PHENIX2010}).  
The influence of {\it event-by-event (e-b-e) fluctuating initial 
conditions} on the photon observables was investigated within the 
(2+1)D Jyv\"askyl\"a ideal hadro model \cite{Chatterjee:2013naa} 
which includes the equilibrated QGP and  Hadron Gas (HG) fluids. 
It has been shown that 'bumpy' initial conditions based on 
the Monte-Carlo Glauber model lead to a slight 
increase at high $p_T$ ($>$ 3 GeV/$c$) for the yield and $v_2$ 
which is, however, not sufficient to explain the experimental data --
see the comparison of model calculations with the PHENIX data 
in Figs. 7,8 of \cite{Chatterjee:2013naa} and with the ALICE data 
in Figs. 9,10 of \cite{Chatterjee:2013naa}.

II.) The influence of {\it viscous corrections} on photon spectra and 
anisotropic flow coefficients $v_n$ has been investigated in two 
viscous hydro models: 
(3+1)D MUSIC  \cite{Dion:2011vd} which is based on 'bumpy'
e-b-e fluctuating initial conditions from impact parameter dependent
Glasma type ('IP-Glasma') and (2+1)D VISH2+1 \cite{Shen13} with 'bumpy'
e-b-e fluctuating initial conditions from the Monte-Carlo Glauber model.
Both hydros include viscous QGP (with lQCD EoS) and HG fluids
and reproduce well the hadronic 'bulk' observables. 
The photon rate has been modified in \cite{Dion:2011vd,Shen13}
in order to account for first order non-equilibrium (viscous) corrections  
to the standard equilibrium rates (i.e. the thermal QGP \cite{AMY01} and 
HG \cite{TRG04} rates).
It has been found that the viscous corrections only slightly
increase the high $p_T$ spectra compared to the ideal hydro calculations 
while they have a large effect on the anisotropic flow coefficients $v_n$
 - see Fig. 9 in  \cite{Dion:2011vd}. 
Interesting that the viscous suppression of hydrodynamic 
flow anisotropies for photons is much stronger than for hadrons. 
Also the photon $v_n$ are more sensitive 
to the QGP shear viscosity which serve the photon $v_n$
as a QGP viscometer \cite{Shen13}.

\begin{figure}[t]
\begin{center}
\begin{minipage}[l]{12cm}
\phantom{a}\vspace*{-1.6cm}\hspace*{-0.5cm}
\includegraphics*[width=11.5cm]{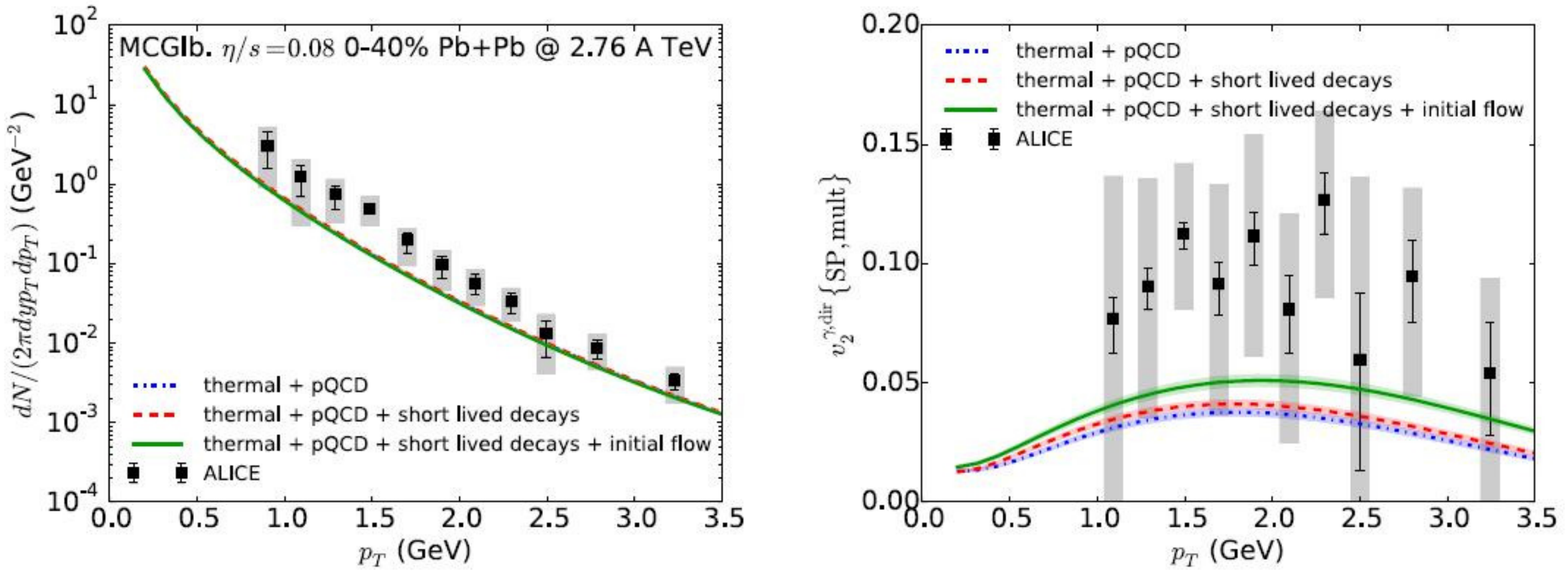}
\end{minipage}
\begin{minipage}[r]{4cm}
\phantom{a}\hspace*{2mm}\vspace*{-4cm}
\caption{Direct photon $p_T$-spectrum and elliptic flow coefficient $v_2$ compared 
with ALICE measurements in $\sqrt{s}=2.76$ TeV Pb+Pb collisions at 0-40\% centrality
\cite{ALICE_v2}.
The figure is taken from Ref. \cite{ShenQM14}.}
\label{fig:initflow}
\end{minipage}
\end{center}
\phantom{a}\vspace*{-4.1cm} \phantom{a}
\end{figure}

III.) Another idea, which has been checked recently within the (2+1)D VISH2+1 
viscous hydro model \cite{Shen13}, is associated with the generation of 
{\it 'pre-equilibrium' flow} (see \cite{ShenQM14}).
The idea of 'initial' flow has been suggested in Ref. \cite{Rapp_inflow}
and modeled as a rapid increase of bulk $v_2$ in the expanding fireball model
which leads to a substantial enhancement of photon $v_2$. 
In a viscous hydro \cite{ShenQM14} the generation of pre-equilibrium flow
has been realized using a free-streaming model to evolve the partons 
to 0.6 $fm/c$ where the Landau matching takes over to switch to viscous hydro.
Such a scenario leads to a quick development of momentum anisotropy with 
saturation near the critical temperature $T_C$ - see Fig. \ref{fig:initflow},
which shows the direct photon $p_T$-spectrum (left) and photon $v_2$ (right)
for  40\% central Pb+Pb collisions at LHC energies in comparison
to the ALICE data \cite{ALICE_v2}. One sees that the pre-equilibrium flow 
effect increases the photon $v_2$ slightly but not sufficient 
to reproduce the ALICE data (the same holds for the PHENIX data).
However, the physical origin of 'initial' flow has to be justified/found
firstly before robust conclusions can be drown.

\subsubsection{Photons from non-equilibrium transport models}

In this Section we depart from the hydro models
and consider the influence of {\it non-equilibrium dynamics}
on photon production. As a 'laboratory' for that 
we will use the microscopic Parton-Hadron-String Dynamics (PHSD)
transport approach \cite{PHSD}, 
which is based on the generalized off-shell 
transport equations derived in first order gradient expansion of the 
Kadanoff-Baym equations, applicable for strongly interacting system.
The approach consistently describes the full evolution of a relativistic 
heavy-ion collision from the initial hard scatterings and string formation 
through the dynamical deconfinement phase transition to the strongly-interacting
quark-gluon plasma  as well as dynamical hadronization and the
subsequent interactions in the expanding hadronic phase as in
the HSD transport approach \cite{CBRep98}.
The partonic dynamics is based on the Dynamical Quasi-Particle Model (DQPM), 
that is constructed to reproduce lattice QCD (lQCD) results for a quark-gluon 
plasma in thermodynamic equilibrium.
The DQPM provides the mean fields for gluons/quarks and their effective 
2-body interactions that are implemented in the PHSD 
(for the details see Ref.~\cite{Cassing:2008nn} 
and \cite{Linnyk:Photon,PHSD}). 
Since the QGP radiation in PHSD occurs from the massive off-shell 
quasi-particles with spectral functions, the corresponding QGP rate 
has been extended beyond the standard pQCD  rate \cite{AMY01} - see Ref. \cite{Linnyk11}.

\begin{figure}[t]
\begin{center}
\phantom{a}\vspace*{-3mm}\hspace*{-1.cm}
\begin{minipage}[l]{6cm}
\phantom{a}\hspace*{-0.8cm}
\includegraphics*[width=8.cm]{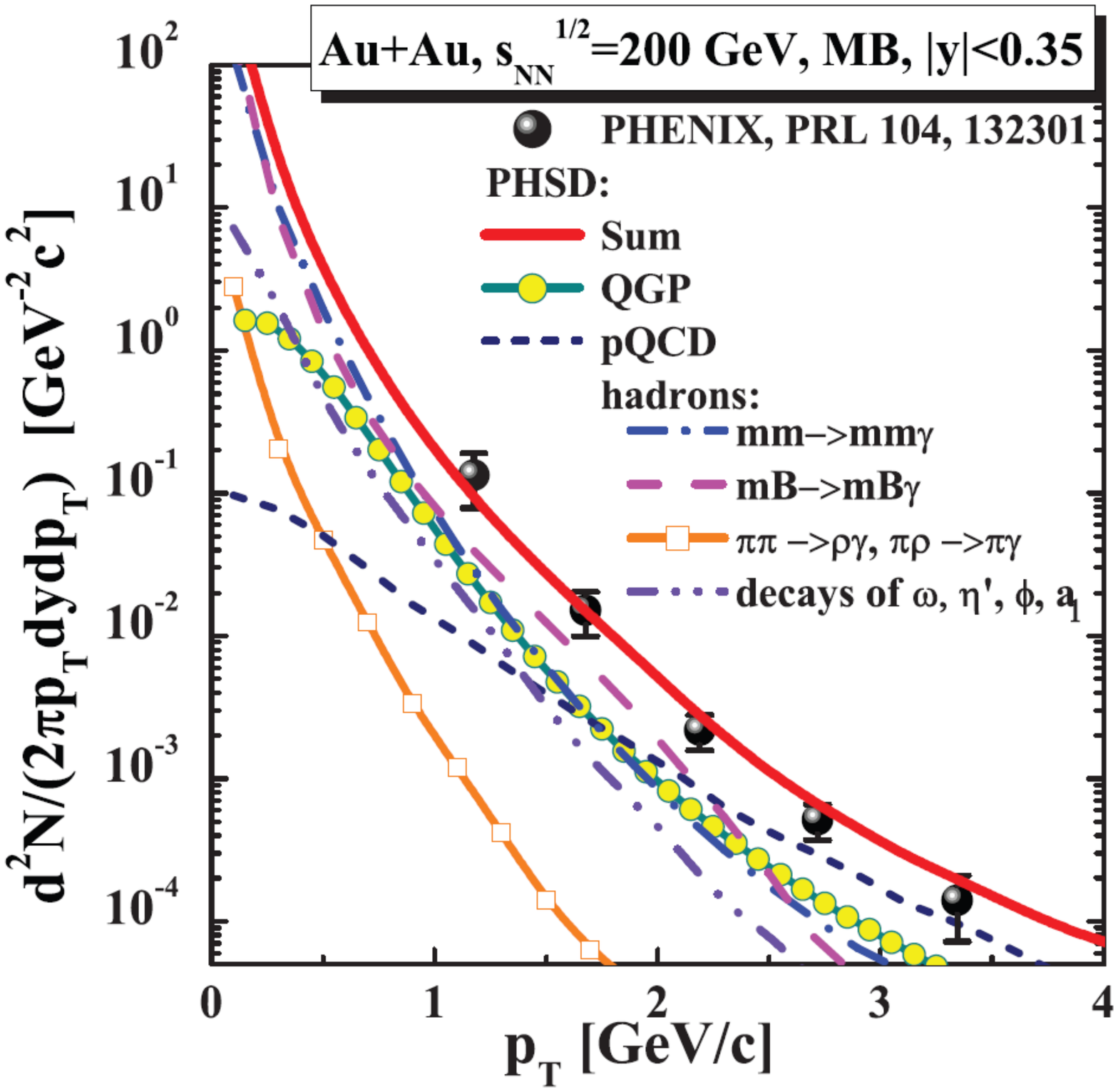}
\phantom{a}\hspace*{0.2cm}
 \begin{minipage}[l]{6.5cm}
\caption{Left: Direct photon $p_T$-spectrum from the PHSD approach 
in comparison to the PHENIX data \cite{PHENIX1} for midrapidity 
minimal bias Au+Au collisions at $\sqrt{s}=200$ GeV.
The figure is taken from Ref. \cite{Linnyk:Photon}.
Right: Centrality dependence of the direct photon $p_T$-spectra  
for 0-20\%, 20-40\%, 40-60\%, 60-92\% central Au+Au collisions 
at $\sqrt{s}=200$ GeV: model predictions vs. the PHENIX data
 \cite{PHENIXcd14}.
The figure is taken from Ref. \cite{MizunoQM14}.}
\label{fig:N3PHSD}
  \end{minipage}
\end{minipage}
\begin{minipage}[l]{10cm}
\phantom{a}\hspace*{0.5cm}
\includegraphics*[width=14.cm,height=9cm]{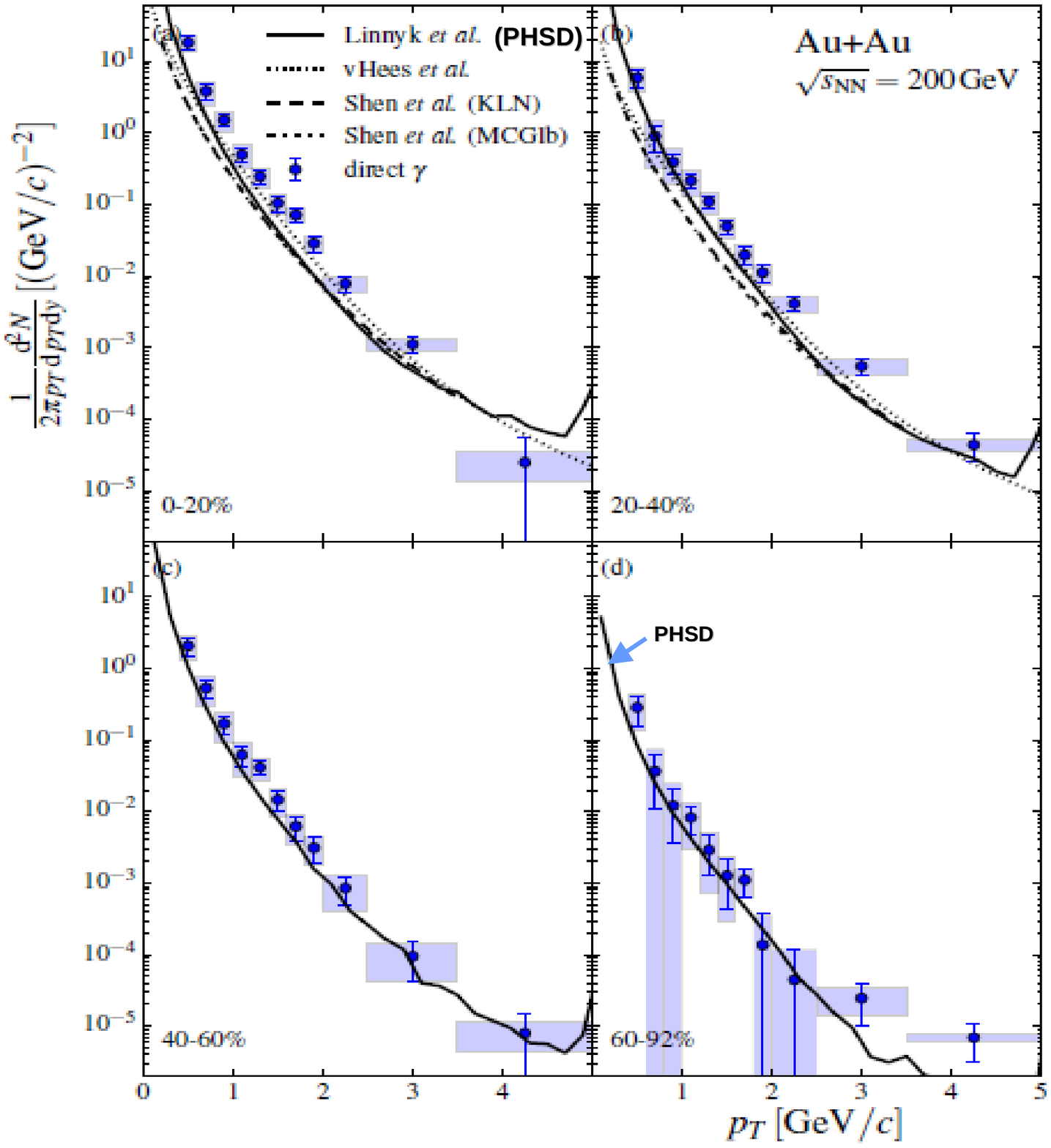}
\end{minipage}
\end{center}
\phantom{a}\vspace*{-1.4cm} \phantom{a}
\end{figure}

The result for the direct photon $p_T$-spectra  from the PHSD approach 
\cite{Linnyk:Photon} is shown in Fig. \ref{fig:N3PHSD} (left)  
in comparison to the PHENIX data \cite{PHENIX1}
for midrapidity minimal bias Au+Au collisions at $\sqrt{s}=200$ GeV. 
While the 'hard' $p_T$ spectra are dominated by the 'prompt' (pQCD) photons,
the 'soft' spectra are filled by the 'thermal' sources: 
the QGP gives up to $~50\%$ of the direct photon yield below 2 GeV/$c$, 
the contribution from binary $mm$ reactions is of subleading order. 
A sizable contribution stems from hadronic sources such as meson-meson ($mm$)  
and meson-Baryon ($mB$) bremsstrahlung, which can be considered as an 
'upper estimate' due to the uncertainties in the implementation 
of photon bremsstrahlung  based on the 'soft-photon' approximation 
(SPA) \cite{Rate2} which implies the on-shell factorization 
of the amplitude $a+b\to a+b+\gamma$ into the strong and electromagnetic parts 
and  assumptions on elastic $mm$ and $mB$ cross sections 
which are little (or not at all) known experimentally
(see details in \cite{Linnyk:Photon}). 
We stress, that $mm$ and $mB$ bremsstrahlung  can not be subtracted 
experimentally from the photon spectra and, thus, has to be included 
in theoretical considerations which is not presently the case for hydro results
presented above where the 'HG' rate isbased on binary mesonic channels 
$m_1+m_2\to m_3+\gamma$ \cite{TRG04}. 
As has been mentioned earlier the importance of bremsstrahlung 
for 'soft' photons follows also from the WA98 data 
at $\sqrt{s}=17.3$ GeV \cite{HSD08,LiuRapp07}, however, more work has 
to be done to provide  robust results on this 'trivial' hadronic source.

\subsubsection{Photon sources: QGP vs. HG? Constraints from data.}

The question: "what dominates the photon spectra - {\it QGP radiation or hadronic sources?}" can be addressed
experimentally by investigating the centrality dependence of the photon
yield: the QGP contribution is expected to decrease from
central to peripheral collisions where the hadronic channels are
dominant. The right part of Fig. \ref{fig:N3PHSD} shows the centrality
dependence of the direct photon $p_T$-spectra  for 0-20\%, 20-40\%, 40-60\%, 60-92\% central Au+Au collisions at $\sqrt{s}=200$ GeV .
The solid dots stand for the recent PHENIX data \cite{PHENIXcd14,MizunoQM14}
whereas the lines indicate the model predictions: solid line - PHSD 
(denoted as 'Linnyk et al.') \cite{Linnyk:Photon}, dashed and dashed-dotted 
lines ('Shen et al. (KLN)' and 'Shen et al.' (MCGib)') are the results
from viscous (2+1)D VISH2+1 \cite{Shen13} and (3+1)D MUSIC  \cite{Dion:2011vd}
hydro models whereas the dotted line ('vHees et al.') stands for the results
of the expanding fireball model \cite{Rapp_inflow}.
As seen from Fig. \ref{fig:N3PHSD} for the central collisions the models deviate 
up to a factor of 2 from the data and each other due to the different 
dynamics and included sources,  for the 
(semi-)peripheral collisions the PHSD results - dominated by $mm$ and $mB$
bremsstrahlung - are consistent with the data which favors these hadronic sources.

The centrality dependence of the direct photon yield, integrated
over different $p_T$ ranges, has been measured by the PHENIX Collaboration, too  \cite{PHENIXcd14,MizunoQM14}. It has been found that the midrapidity 'thermal' photon yield scales with the number of participants as 
$dN/dy \sim N_{part}^\alpha$ with $\alpha =1.48\pm 0.08$
and only very slightly depends on the selected $p_T$ range (which is 
still in the 'soft' sector, i.e. $< 1.4$ GeV/$c$).
Note that the 'prompt' photon contribution (which scales 
as the $pp$ 'prompt' yield times the number of binary collisions in $A+A$) 
has been subtracted from the data. The PHSD predictions \cite{Linnyk:Photon}
for the minimal bias Au+Au collisions give $\alpha (total) =1.5$ which is 
dominated by hadronic contributions while the QGP channels scale with 
$\alpha (QGP) \sim 1.7$. \ 
A similar finding has been obtained by the viscous (2+1)D VISH2+1 
and (3+1)D MUSIC hydro models \cite{VISH_McG}: 
$\alpha(HG) \sim 1.46, \ \  \alpha(QGP) \sim 2, \ \ \alpha(total) \sim 1.7$.
Thus, the QGP photons show a centrality dependence significantly 
stronger than that of HG photons. 

Another question: "can one use the photon spectra  as a {\it 'thermometer' 
of the QGP?}", i.e. to obtain the temperature of photon sources by extracting 
the slope parameter from a thermal fit of the photon yields. This question 
has been answered in a detailed study \cite{VISH_McG} by the viscous 
(2+1)D VISH2+1 and (3+1)D MUSIC hydro models where it has been shown that 
the measured $T_{eff} > $ 'true' $T$ due to the 'blue shift' induced  by 
the strong radial flow. Moreover, it has been shown that only 
$\sim 1/3$ at LHC and $\sim 1/4$ at RHIC of the total photons come
from the 'hot' QCD with $T>250$~MeV.

\begin{figure}[t]
\begin{center}
\begin{minipage}[l]{12.5cm}
\hspace*{-0.5cm}
\includegraphics*[width=6.8cm]{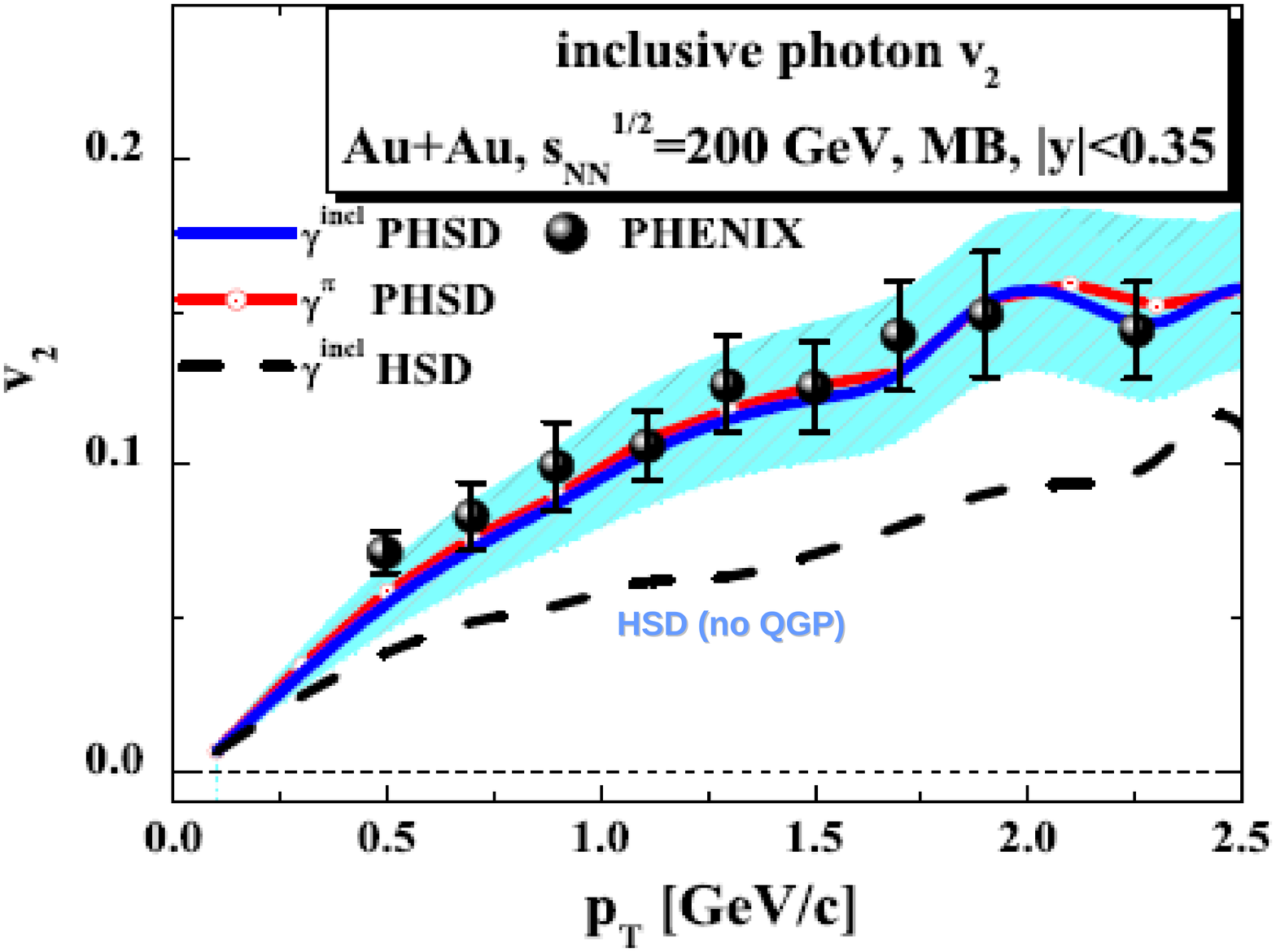}
\includegraphics*[width=6.8cm]{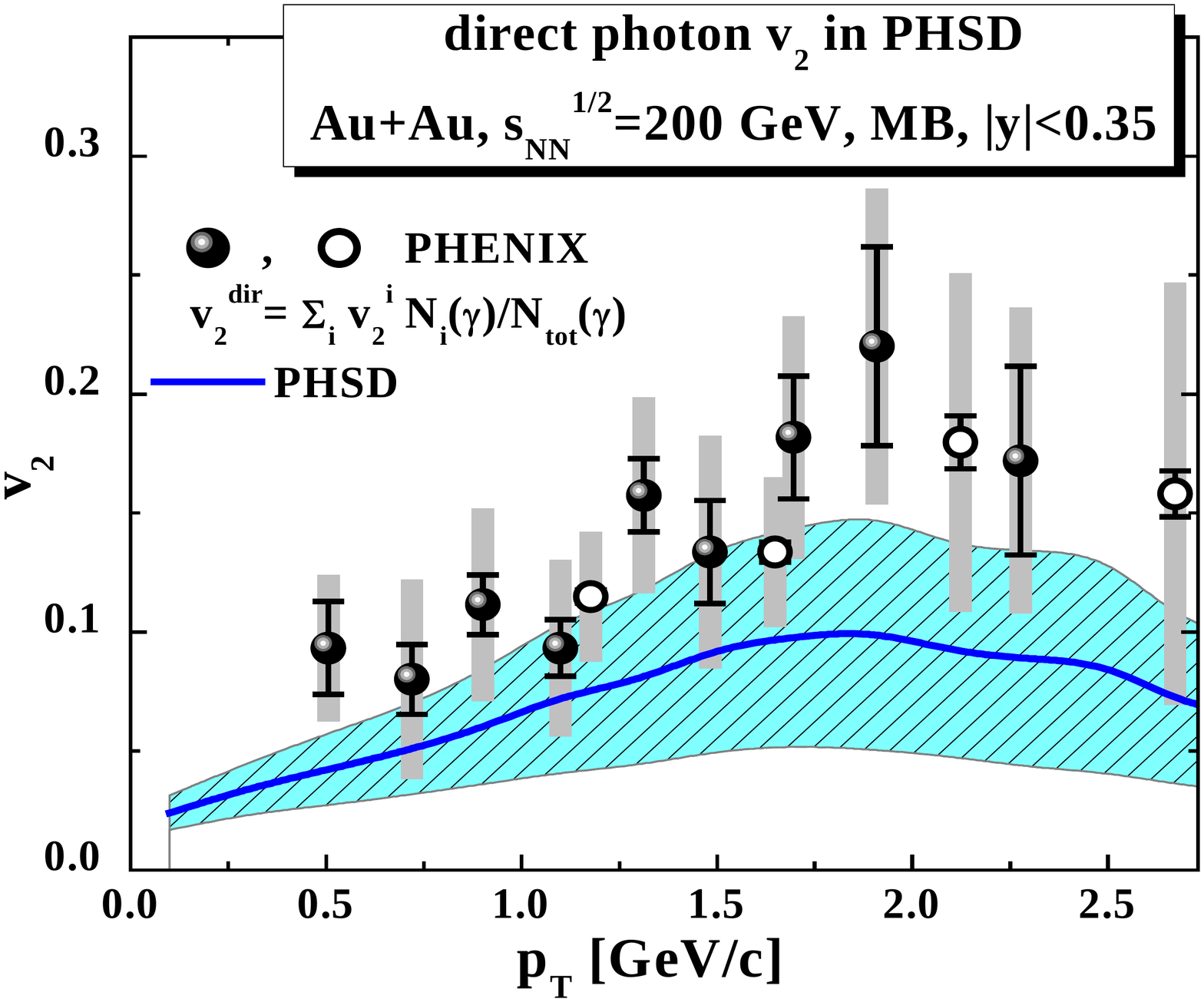}
\end{minipage} 
\begin{minipage}[r]{3.5cm}
\vspace*{-0.3cm}
\caption{Inclusive (left) and direct (right) photon  elliptic flow coefficient
$v_2(p_T)$ from the PHSD approach in comparison to the PHENIX data \cite{PHENIX1} 
for midrapidity minimal bias Au+Au collisions at $\sqrt{s}=200$ GeV.
The figure is taken from Ref. \cite{Linnyk:Photon}.}
\label{fig:v2PHSD}
\end{minipage}
\end{center}
\phantom{a}\vspace*{-1.5cm} \phantom{a}
\end{figure}

Do we see finally the QGP pressure in $v_2$ if the photon production is
dominated by hadronic sources or are the direct photons a {\it 'barometer' 
of the QGP?} As argued in \cite{Linnyk:Photon} and illustrated in
Fig. \ref{fig:v2PHSD}, one can answer the question positively since 
the QGP causes the strong elliptic flow of photons indirectly, 
by enhancing the $v_2$ of final hadrons due to the partonic interactions.
Due to that in PHSD calculations the $v_2$ of inclusive photons from 
minimal bias midrapidity Au+Au collisions at $\sqrt{s}=200$ GeV, which 
are mainly coming from $\pi_0$ decays (blue line in the left part of 
Fig. \ref{fig:v2PHSD}), is similar to 
the pion $v_2$ (red line) and consistent with the PHENIX data, while the 
HSD results, i.e. without QGP, underestimates the $v_2$ of inclusive photons 
(as well as hadrons) by a factor of 2 (dashed line).   
However, the direct photon $v_2$ puzzle remains in the PHSD (right part of 
Fig. \ref{fig:v2PHSD}) due to the very small $v_2(\gamma^{QGP})$
which contributes is up to 50\% of the total yield. The error band shows 
the estimated uncertainties related to the $mm$ and $mB$ bremsstrahlung 
which dominate the direct photon $v_2$ in PHSD.  

We note that there are other scenarios towards the solution of the direct photon $v_2$ puzzle discussed during the 'Quark Matter-2014' such as
early-time magnetic field effects \cite{Magnet},
Glasma effects \cite{Larry}, 	
pseudo-critical enhancement of thermal photons near $T_C$  \cite{Rapp0414}
or non-perturbative effects of 'semi-QGP' \cite{LinQM14}.

\section{Dileptons}

\subsection{Dilepton production sources}

Dileptons ($e^+e^-$ or $\mu^+\mu^-$ pairs) can be emitted from all 
stages of the reactions as well as photons. One of the advantages of 
dileptons compared to photons is an additional 'degree of freedom' 
- the invariant mass $M$ which allows to disentangle various sources.\\
1) Hadronic sources of dileptons in $p+p, p+A$ and $A+A$ collisions:\\
$(i)$ at low invariant masses ($M < 1$ GeV$c$) -- the  Dalitz decays of mesons and
 baryons $(\pi^0,\eta,\Delta, ...)$ and the direct decay of  
vector mesons  $(\rho, \omega, \phi)$ as well as hadronic bremsstrahlung; \\
$(ii)$ at intermediate masses ($1< M < 3$ GeV$c$) -- 
leptons from correlated $D+\bar D$ pairs, radiation 
from multi-meson reactions 
($\pi+\pi, \ \pi+\rho, \ \pi+\omega, \ \rho+\rho, \ \pi+a_1, ... $)  - 
so called $'4\pi'$ contributions; \\
$(iii)$ at high invariant masses ($M > 3$ GeV$c$) -- the direct decay of 
vector mesons  $(J/\Psi, \Psi^\prime)$ and
initial 'hard' Drell-Yan annihilation to dileptons
($q+\bar q \to l^+ +l^-$, where $l=e,\mu$).\\
2) 'Thermal' QGP dileptons radiated from the partonic interactions 
in heavy-ion $A+A$ collisions that contribute dominantly to 
the intermediate masses. The leading processes
are the 'thermal' $q\bar q$ annihilation ($q+\bar q \to l^+ +l^-$, \ \
$q+\bar q \to g+ l^+ +l^-$) and Compton scattering 
($q(\bar q) + g \to q(\bar q) + l^+ +l^-$). 

\subsection{Dileptons: from SPS to LHC energies}

Dileptons from heavy-ion collisions have been measured in the last decades 
by the CERES \cite{CERES} and NA60 \cite{NA60} Collaborations at SPS energies. 
The high accuracy dimuon NA60 data provide a unique possibility 
to subtract the hadronic cocktail from the spectra
and to distinguish  different in-medium scenarios for the $\rho$-meson
spectral function such as a collisional broadening and dropping mass 
\cite{ChSym,Li:1995qm}.
The main messages obtained by a comparison of the variety of model
calculations (see e.g. \cite{ChSym,dilHSD,dilSPStheor}) with 
experimental data can be summarized as  
(i) low mass spectra \cite{CERES,NA60} provide a clear evidence 
for the collisional broadening of the $\rho$-meson spectral function
in the hot and dense medium;
(ii)~intermediate mass spectra above $M>1$ GeV/$c^2$ \cite{NA60}
are dominated by partonic radiation;
(iii) the rise and fall of the inverse slope parameter of the 
dilepton $p_T$-spectra  (effective temperature) $T_{eff}$ 
\cite{NA60} provide evidence for thermal QGP radiation;
(iv) isotropic angular distributions \cite{NA60} are an indication 
for a thermal origin of dimuons.

An increase in energy from SPS to RHIC has opened  new possibilities
to probe (by dileptons) the matter at very high temperature, i.e.  
dominantly in the QGP stage, created in central heavy-ion collisions.
The dileptons ($e^+e^-$ pairs) have been measured by 
the PHENIX Collaboration 
for $pp$ and $Au+Au$ collisions at $\sqrt{s}=200$ GeV \cite{PHENIXdil}. 
A large enhancement of the dilepton yield relative to the scaled $pp$ 
collisions in the invariant mass regime from 0.15 to 0.6 GeV/$c^2$ 
has been observed. This observation has stimulated a lot of theoretical
activity (see the model comparison with the data in \cite{PHENIXdil}).
The main messages - which stay up-to-now - can be condensed such that
the theoretical models, providing a good description of $pp$ dilepton data 
and peripheral $Au+Au$ data, fail in describing the excess in 
central collisions even with in-medium scenarios for the vector 
meson spectral function. 
The missing strengths might be attributed to low $p_T$ sources  
\cite{dilPHSDRHIC}.
The intermediate mass spectra are dominated by the QGP radiation as well as
leptons from correlated charm pairs ($D+\bar D$) 
 \cite{dilHSD,dilPHSDRHIC,Rapp13}.

\begin{figure}[t]
\begin{center}
\phantom{a}\vspace*{-0.4cm}
 \begin{minipage}[l]{11cm}
\includegraphics*[width=10.5cm]{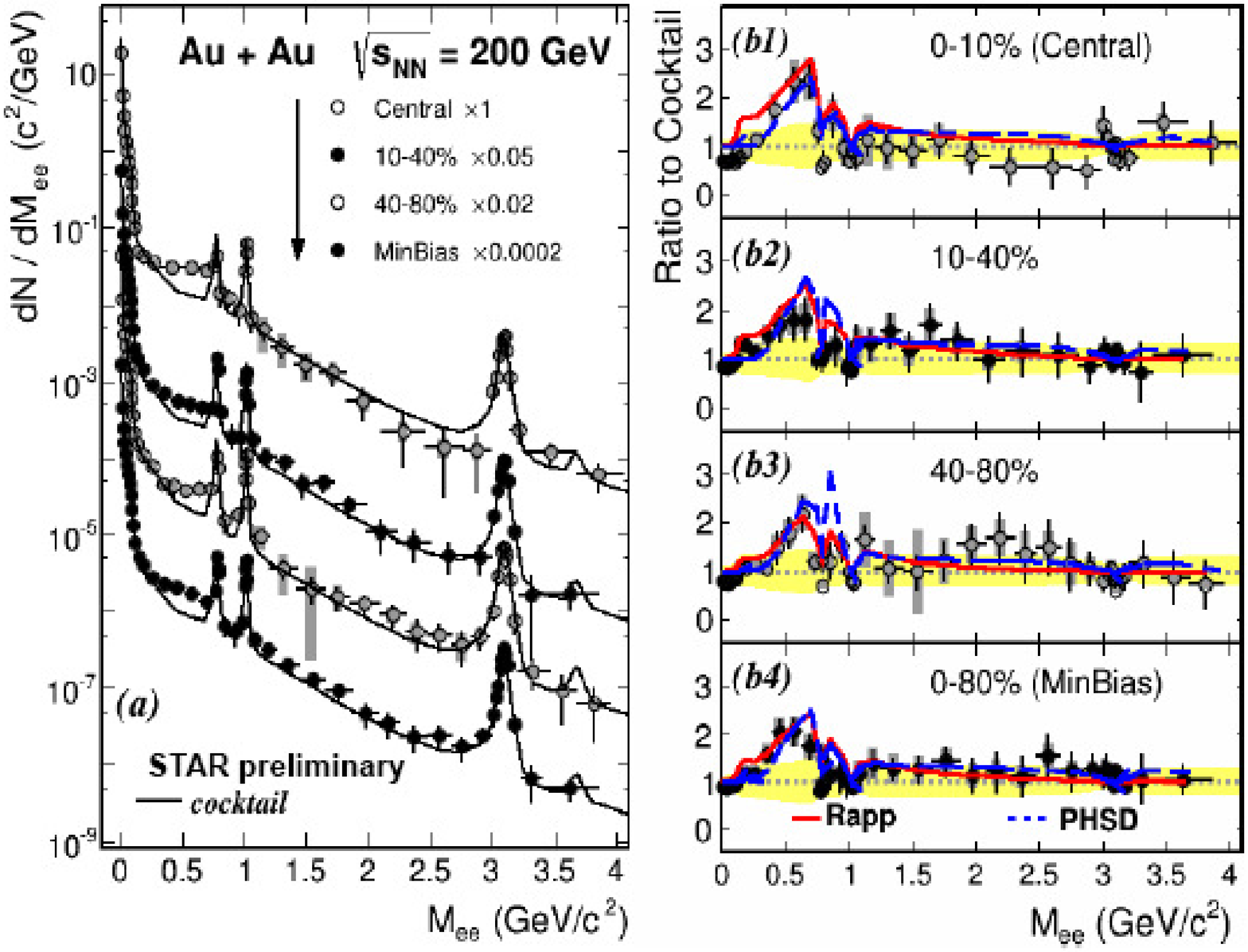}
\end{minipage}
 \begin{minipage}[r]{5cm}
\caption{Centrality dependence of the midrapidity dilepton yields (left)
and its ratios (right) to the 'cocktail' for 
0-10\%, 10-40\%, 40-80\%, 0-80\% central Au+Au collisions at $\sqrt{s}=200$ GeV:
a comparison of STAR data with theoretical predictions from the PHSD 
('PHSD' - dashed lines) and the expanding fireball model ('Rapp' - solid lines).
The figure is taken from Ref. \cite{HuckQM14}.}
\label{fig:dilSTAR200cd}
\end{minipage}
\end{center}
\phantom{a}\vspace*{-1.7cm} \phantom{a}
\end{figure}

In this respect it is very important to have independent measurements 
which have been carried out by the STAR Collaboration \cite{dilSTAR}.
Fig. \ref{fig:dilSTAR200cd} shows the comparison of STAR data 
of midrapidity dilepton yields (left) and its ratios (right) 
to the 'cocktail' for  0-10\%, 10-40\%, 40-80\%, 0-80\% central 
Au+Au collisions at $\sqrt{s}=200$ GeV  in comparison to 
the theoretical model predictions from the PHSD  and the 
expanding fireball model. 
As seen from Fig. \ref{fig:dilSTAR200cd} the excess of the dilepton yield 
over the expected cocktail is larger for very central collisions and consistent 
with the model predictions based on the collisional broadening of 
the $\rho$-meson spectral function  and QGP dominated radiations at
intermediate masses.
Moreover, the recent STAR dilepton data for Au+Au collisions 
from the Beam Energy Scan (BES) program  for the $\sqrt{s}=19.6, 27, 39$ and 
62.4 GeV \cite{RuanQM14,HuckQM14} are also in line with 
the expanding fireball model predictions with a $\rho$ collisional broadening
\cite{HuckQM14}. According to the PHSD predictions,
the excess is increasing with decreasing energy due to a longer 
$\rho$-propagation in the high baryon density phase (see Fig. 3 in \cite{RuanQM14}).

The upcoming PHENIX data for central Au+Au collisions
- obtained after the upgrade of the detector -
together with the BES-II RHIC data should provide finally a consistent picture
on the low mass dilepton excess. 
The future ALICE data \cite{dilALICE} 
for heavy-ions will give a clean access to the dileptons emitted 
from the QGP \cite{Rapp13,dilPHSDLHC}.

Last but not least, we mention that promising perspectives with dileptons 
have been suggested in Ref. \cite{v3dil} - to measure the anisotopic
coefficients  $v_n, \ n=2,3$ similar to photons. The calculations 
done with the viscous (3+1)d MUSIC hydro for central Au+Au collisions 
at RHIC energies show that $v_2, v_3$ are sensitive to the dilepton sources and
to the EoS and $\eta/s$ ratio. The main advantage of measuring 
$v_n$ with dileptons compared to photons is the fact that an extra degree 
of freedom $M$ allows to disentangle the sources. However, this is a very 
challenging experimental task. 

\section{Conclusions}

I. The main messages from the 'photon adventure' can be summarize in short as:
$(i)$ the photons  provide a critical test for the theoretical models:
the standard dynamical models - constructed to reproduce the 'hadronic world' - 
fail to explain the photon experimental data; 
$(ii)$ The details of the hydro models (fluctuating initial conditions, 
viscousity, pre-equilibrium flow) have a small impact on the photon observables; 
$(iii)$ as suggested by the PHSD transport model calculations 
the role of trivial sources such as $mm$ and $mB$ bremsstrahlung has been underestimated and has to be re-considered; 
$(iv)$ The initial phases of the reaction might turn out to be important
(pre-equilibrium /'initial' flow, Glasma effect etc.).
Finally one may conclude that the photons are one of the most sensitive probes for the dynamics of HIC and for the role of the partonic phase.

II. The main messages from dilepton data are
$(i)$ at low masses ($0.2-0.6$ GeV/$c^2$) dilepton spectra show sizable 
changes due to hadronic in-medium effects, i.e. modification of the 
properties of vector mesons (such as collisional broadening) in the hot 
and dense hadronic medium, related to the chiral symmetry restoration;
these effects can be observed at all energies from SIS to LHC;
$(ii)$ at intermediate masses: the QGP ($q\bar q$ thermal radiation)  
dominates for $M>1.2$ GeV/$c^2$. 
The fraction of QGP grows with increasing energy and becomes dominant  
at the LHC energies. 

Finally, the dilepton measurements within the future experimental energy and system 
scan ($pp, pA, AA$) from low to top RHIC energies as well as 
the new ALICE data at LHC energies will extend our knowledge on the properties of hadronic and partonic matter.

\vspace*{1mm} 
The author acknowledges the financial support through the 'HIC for FAIR' framework of the 'LOEWE' program.







\phantom{a} 

\end{document}